 \documentclass [12pt,a4paper      ]{article}
\usepackage{times}

\DeclareFontFamily{OT1}{times}{}
\DeclareFontShape {OT1}{times}{m }{n }{ <-> ptmr }{}
\DeclareFontShape {OT1}{times}{bx}{n }{ <-> ptmb }{}
\DeclareFontShape {OT1}{times}{m }{it}{ <-> ptmri}{}
\DeclareFontShape {OT1}{times}{bx}{it}{ <-> ptmbi}{}
\usepackage{amsmath}
\usepackage{amsfonts}
\usepackage{amssymb}
\usepackage{latexsym}
\numberwithin{equation}{section}
\newcommand{\cl}{C \kern -0.1em \ell} 

\newcommand{\VEC}{\vec{\kern +.1em[}} 
\newcommand{\TOR}{\vec{\kern +.2em]}} 
\newcommand{\BRA}{\langle\kern -.2em\langle} 
\newcommand{\KET}{\rangle\kern -.2em\rangle} 

\begin{document}

\title{\bf\vspace{0.0cm} 
       More on the early interpretation of the Schwarzschild solution}

\author{Andre Gsponer}

\date{ISRI-04-09.3 ~~ \today}

\maketitle

\begin{abstract}

G.\ Lema\^itre was apparently the first to make an explicit coordinate transformation resulting in the removal of the singularity at $r=a=2m$ in the Schwarzschild metric, while C.\ Lanczos was the first to express doubts on the physical reality of that singularity since it could be introduced or removed by a transformation of coordinates.

\end{abstract}


In a recent paper Hans-J\"urgen Schmidt \cite{SCHMI2004-} emphasized the important contribution made by J.L.\ Synge in 1950 to the clarification of the physical interpretation of the Schwarzschild black hole at its horizon \cite{SYNGE1950-}.

The paper of Synge is indeed remarkable, especially because it ``contains many interesting results all of them being discussed in details'' \cite{HECKM1951-}.  What is also remarkable is that Synge's paper is greatly inspired by the work of G.\ Lema\^itre, who apparently was the first to have made an explicit  transformation of coordinates to remove the singularity at $r=a=2m$ in the Schwarzschild metric \cite{LEMAI1933-}.  Moreover, Synge mentions H.P. Robertson's lectures in 1939 at the University of Toronto, which drew his attention to Lema\^itre's paper, as well as to similar results obtained by H.P. Robertson.

   Therefore, while Synge's paper is certainly an outstanding achievement, it is not the first in which the Schwarzschild horizon was correctly interpreted.  In fact, according to Jean Eisenstaedt \cite{EISEN1989-}:
\begin{quote}
``The first expert to cast doubt on the reality of the singular character [of the Schwarzschild solution] at $r=2m$ was Cornelius Lanczos~\cite{LANCZ1923-}. [...] Lanczos introduced a new ``singularity'' [i.e., the Schwarzschild singularity] at a place where the metric was completely regular before, thus showing the relativity of the then vague concept of a singularity.''
\end{quote}

   Hence, by introducing rather than removing a singular term, Lanczos did the ``opposite'' of what Lema\^itre was apparently the first to have done.  Furthermore, generalizing from the case of the Schwarzschild metric, Lanczos concluded his paper by stressing that \cite[p.539]{LANCZ1923-}:
\begin{quote}
 ``this example proves how little one might conclude from a singular behavior of the $g_{ij}$ functions about a real singularity of the field, for it can possibly be eliminated by a coordinate transformation.''
\end{quote}

   This paper by Lanczos was his first publication after his doctoral dissertation: a field theoretical interpretation of electrodynamics in which singularities of the homogeneous Maxwell's equations are interpreted as electrons \cite{LANCZ1919-, HURNI2004-}.  This may explain why Lanczos paied so much attention to singularities in Einstein's field, and consequently published several seminal papers related to singularities, surface distributions of matter, and junction conditions in general relativity  \cite{DAVIS1998-}.

\end{document}